\documentclass[aps,amsmath,twocolumn,showpacs]{revtex4}
\usepackage{graphicx}
\begin{document}

\preprint{APS/123-QED}

\title{Photothermally-Induced Nonlinearity in a Quantum Multimode Optical System}
\author{Akhtar Munir}
\email{amkhan@mail.ustc.edu.cn}
\affiliation{Department of Physics, Zhejiang Normal University, Jinhua, 321004, China}
\date{\today}
\begin{abstract}
Similar to radiation
pressure, photothermal effects connect the optical path length to an intracavity field, resulting in nonlinear behavior of the resonator due to thermal effects. Here, we theoretically investigate the nonlinear optics that emerge as a result of photothermal effects in a multimode optical system composed of two cavity modes coupled via hopping coefficient and with two mechanical modes through coupling rates. We report single to double photothermally-induced transparency (PTIT) dips, and a single sharp photothermally-induced absorption (PTIA) peak, and demonstrate that photothermal and strong coupling coefficients can suppress this phenomenon. Moreover, we observe Fano resonances in the absorption profile by monitoring probe transmission in the off-resonant configuration of the transparency phenomenon. The dynamics of group delay or advance are investigated in the range of transparency such that a sharp dip can assist in achieving slow light for a longer time. Using appropriate experimental parameters, our proposed work can pave the way for future practical applications in quantum information processing based on multimode interactions.
  
\end{abstract}

\date{\today}

\maketitle
\section{Introduction}
\label{Sec:intro}

Various phenomena related to radiation pressure has been investigated for many years within the framework of mechanically compatible optical cavities for damping of micromechanical dynamics~\cite{aspelmeyer2014,favero1}, actuation~\cite{faver2} and sensing~\cite{vahala2003}. A variety of optical cavity designs are utilized for this purpose, including Fabry-Perot cavities with movable end mirrors, whispering-gallery glass microtoroids, and nanoscale guided-wave devices. The interaction of optical beams with the mechanical oscillation of a mirror in an optical cavity via radiation pressure assists in the investigation of various physical phenomena, including optomechanically induced transparency (OMIT)~\cite{Agarwal2010,weis2010,verhagen2012}, squeezing of light~\cite{purdy2013}, modulation of light~\cite{xiong2017,xiong2016}, quantum knowledge~\cite{stannigel201,komar2013}, precision measurement~\cite{zhang2012,wang2015}, and heavy coupling physics~\cite{groblacher2009}.

In addition to radiation pressure effects, photothermal effects also provide highly effective optomechanical interaction between cavity photons and the mechanical degrees of freedom~\cite{restrepo2011}. In photothermal effects, the mechanical oscillator absorbs cavity photons, causing thermo-elastic distortion and displacement of the oscillator. Depending on the interaction, these photothermal effects may reduce or extend the cavity's optical path~\cite{konthasinghe2017,an1997,munir2021}. Since photothermal pressure is several orders of magnitude higher than radiation pressure~\cite{metzger2008}, it is of great interest to investigate photothermal effects in various optical systems. On the one hand, it has the potential to reduce the accuracy of high-sensitivity interferometer displacement measurements~\cite{braginsky1999,evans2008}, from microscopic cantilever optical cavities~\cite{kleckner2006} to kilometer-scale gravitational-wave detectors such as LIGO~\cite{braginsky1999,abbott2016,harry2006}. In extreme situations, the shot noise of absorbed light could place a basic limit on measuring sensitivity~\cite{ludlow2007,cerdonio2001} and the production of low-frequency squeezing~\cite{goda2005}. Photothermal effects, on the other hand, have been demonstrated to be useful in decreasing mechanical oscillator Brownian noise~\cite{metzger2004c,hosseini2014} and may even be capable of cooling a mechanical resonator close to its quantum ground state~\cite{pinard2008}.

Recently, there has been a growing interest in multimode interactions in optical systems~\cite{gao2015,riedinger2018,lei2015three}, which are common in quantum physics and have been investigated in a variety of physical systems. For example, phonon-photon-phonon~\cite{lin2010,massel2012,spethmann2016} and photon-phonon-photon~\cite{tian2013,dong2012,wang2012,wang2013} are two common multimode interactions in optomechanical systems. Until now, radiation pressure has only been explored mainly for the purpose of investigating multimode interactions in optomechanical systems. For example, fast and robust quantum control~\cite{zhang2019}, three-pathway electromagnetic induced transparency (EIT)~\cite{lei2015three}, and dynamical multistability~\cite{gao2015} play an important role in justifying the feasibility of multimode interactions in practical applications of quantum information processing~\cite{stannigel201}, multiple pathways interference phenomena, and have applications in memory storage~\cite{massel2012} and insensitive force or displacement detections~\cite{Anetsberger2009}. Thus, it is worth to investigate multimode interactions in an optical system that includes photothermal effects, which, like radiation pressure, connect the optical length to an intracavity field, heating the resonator and causing it to behave nonlinearly.

Here, we aim to investigate the nonlinear optics mediated by photothermal effects in a multimode optical system composed of two cavities and two mechanical modes. On the left and right sides of the cavity, external probe and control laser fields act via a waveguide. We observed single to double photothermally-induced transparency (PTIT) dips, one is sharp, and a single photothermally-induced absorption (PTIA) peak, and noticed that increasing the photothermal and coupling coefficients suppressed the amplitude of the dips. This sharp PTIT dip and PTIA peak are similar to previous work, which observed a sharp EIT dip~\cite{lei2015three} and electromagnetic induced absorption (EIA) peak~\cite{qu2013} via radiation pressure in a coupled-cavity system. 
Here, in the presence of photothermal effects, the PTIT dip is formed by destructive interference between the probe field and anti-Stokes sidebands of the control field, rather than the interference of quantum paths like the EIT or OMIT~\cite{ma2020}. Moreover, we analyze the phenomenon of Fano resonances in the absorption profile that occurs in an off-resonant configuration of PTIT. The asymmetric Fano resonance is a signature of interacting quantum systems, though its spectral shape may vary significantly in different symmetric resonance curves scenarios such as symmetry in EIT windows~\cite{qu2013} or a Lorentzian resonance~\cite{miroshnichenko2010}.
Several schemes have been proposed in literature where group delay and advance can be transformed to each other. However, for appropriate and compact device manufacturing for velocity control analysis, both group delay and advance in the same medium at different frequencies are required. In our proposed scheme, we examine the dynamics of the group delay or advance in the region comprising the two photothermally-induced windows and observe that the sharp PTIT dip provides us with a slow and fast light at two different frequencies simultaneously.

The paper is structured as follows. In Sec.~\ref{Sec:model} we present our theoretical model, the Hamiltonian description of the system, and derive the quantum Langevin equations (QLEs) from the Heisenberg equation of motion. Following the standard input-output relation, a mathematical representation of the outgoing field is constructed, which is then employed to calculate the transmission of the group delay. In Sec.~\ref{Sec:results}, we present numerical findings and discussions about the multimode single to double PTIT, Fano resonances, and the dynamics of the group delay. In Sec.~\ref{Sec:conclusions}, we end with conclusions and suggestions for further work.
\section{Theoretical Model}
\label{Sec:model}
\begin{figure}
\centering
\includegraphics[width=1\linewidth]{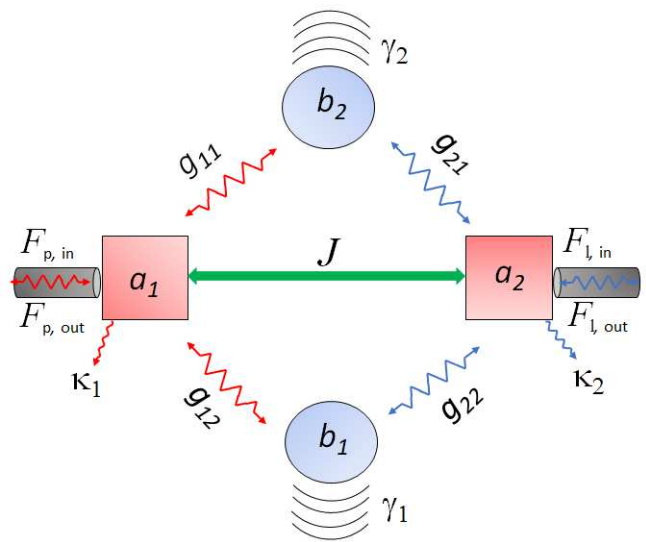}
\caption{Schematic of an optical system composed of two cavity modes, $a_1$ and $a_2$, and two mechanical modes, $b_1$ and $b_2$. The cavity modes are linearly coupled through hopping coefficient $J$, and with mechanical modes via coefficient rates $g_{\imath \jmath}$, for $\imath\in\{1,2\}$ and $\jmath\in\{1,2\}$. Here, $\kappa_1$($\kappa_2$) and $\gamma_1$($\gamma_2$) are the decay and damping rates of the cavity mode $a_1$($a_2$) and mechanical mode $b_1$($b_2$), respectively. The probe ($F_\text{p}$) and control ($F_\text{l}$) fields are applied via waveguides on the left and right sides of the cavity mode, respectively.}
\label{fig:figure01}
\end{figure}

In this section, we present a theoretical model of our system for the purpose of investigating photothermal effects in optomechanical cavity settings. We consider an optical system composed of two microcavity modes $a_1$ and $a_2$ that are linearly coupled with two mechanical modes $b_1$ and $b _2$ (see Fig. \ref{fig:figure01}). The optical modes in the two cavities are connected with each other through the coupling coefficient $J$, while they are coupled with mechanical modes via the coupling constants $g_{\imath\jmath}$.
Moreover, the cavity modes are connected to two external waveguides, which transmit weak probe and strong control field signals along the right and left sides, respectively. In a manner similar to the problem of radiation pressure, photothermal effects couple the cavity optical path length to the intracavity power. This is due to the absorption of photons by the cavity mirrors resulting in thermal expansion and refractive index change of the mirror coating and substrate. These photothermal effects can either decrease or increase the optical path length of the cavity depending on the interaction. Just as with radiation pressure, a modulation of cavity length caused by the photothermal process can lead to nontrivial feedback interrelations between intracavity power and cavity length.
Under the dipole and rotating wave approximations, we can write the total Hamiltonian of the system as   
\begin{align}
\label{eq:ham}
H=&\sum_{\imath=1}^{2}\left[\frac{p_\imath ^2}{2 m _\imath}+\frac{1}{2}m _\imath  \omega_{\rm m_\imath} ^2 x_\imath ^2+\hbar \Delta_\imath a_\imath ^ {\dagger} a_\imath\right]\nonumber \\ -&\sum_{\imath,\jmath=1}^{2}\hbar g_{\imath\jmath}a_\imath ^ {\dagger}a_\imath x_\jmath-\hbar J\left[a_1^ {\dagger}a_2 + a_1 a_2^ {\dagger}\right]\nonumber \\+&\text{i}\hbar\left[F_\text{p} \text{e}^{-\text{i}\Delta_\text{p} t}a_1^{\dagger}-F_\text{p}^{*}\text{e}^{\text{i}\Delta_\text{p} t}a_1\right]+\text{i}\hbar F_\text{l}\left[ a_2^{\dagger}-a_2\right],
\end{align}
where the first term describes the unperturbed part of the Hamiltonian of the mechanical modes and cavity field. In Eq.~(\ref{eq:ham}), $m$ signifies the mass of the mechanical mode, $\omega_{\rm m}$ is the frequency of the mechanical mode, $x$ and $p$ are the position and momentum of the mechanical mode, $a^\dagger$ ($a$) represents the creation (annihilation) operator, respectively, while, $\Delta_\imath=\omega_\imath-\omega_{\text{p},\text{l}}$ defines the detuning of the cavity field frequency and $\Delta_\text{p}=\omega_\text{p}-\omega_\text{l}$ the detuning of the probe field frequency. The second term in the Hamiltonian describes the interaction between cavity modes and mechanical modes with coupling rate $g_{\imath\jmath}$. The third term denotes the interaction of two optical cavities with a coupling coefficient $J$. The last two terms in Eq.~(\ref{eq:ham}) refer to the probe and control fields applied to the left and right optical cavities, respectively. The amplitude of the probe field is given by
\begin{equation}
|F_\text{p}|=\sqrt{\frac{2\kappa_1 P_\text{p}}{\hbar \omega_\text{p}}},
\end{equation}
and amplitude of control field
\begin{equation}
	|F_\text{p}|=\sqrt{\frac{2\kappa_1 P_\text{p}}{\hbar \omega_\text{p}}},
\end{equation}
where $\kappa_{1,2}$ denotes the cavity decay rate, $P_\text{p}$ ($P_\text{l}$) represents the power of probe (control) field, and $\omega_{\text{p},\text{l}}$ is the frequency of the external field.

Next, we employ the Heisenberg equation of motion in order to describe the dynamics of the system. For any generic operator $\mathcal{O}$, the stochastic equation of motion can be expressed as follows:
\begin{equation}
	\label{eq:Eqmn}
\frac{\text{d} \mathcal{O}}{\text{d} t}=-\frac{\text{i}}{\hbar}[\mathcal{O}, H]-\gamma \mathcal{O}+\mathcal{N},
\end{equation}
where $\gamma$ denotes the decay rate associated with the cavity photon and mechanical modes, while $\mathcal{N}$ signifies the Brownian and input vacuum noise operator relevant to the cavity field. Before exploring the system's dynamics, we first address the issue of the photothermal effects caused by cavity photons, which heat up the mirror and cause photothermal displacement. Such changes in optical field displacement cause nonlinearity in optical systems, which we intend to investigate here. We assume that the photothermal displacement is linearly dependent on the temperature around the equilibrium~\cite{marino2006}. From this it follows that the photothermal displacement can be expressed as~\cite{ma2020}
\begin{equation}
\label{eq:heisen}
\dot{x}_{\imath}^{\text{th}}=-\gamma_\imath \left(x_{\imath}^{\text{th}}+ \beta\sum_{\jmath=1}^{2} g_{\imath\jmath} a_{\imath}^{\dagger}a_{\imath}\right),
\end{equation}
$\imath\in \{1,2\}$, where $x^\text{th}$ denotes photothermal displacement, $\gamma_\imath$ represents the effective photothermal relaxation rate, and $\beta$ is the effective photothermal coefficient with unit m/W.

After introducing the photothermal displacement and by inserting the Hamiltonian~(\ref{eq:ham}) into the Heisenberg equation of motion~(\ref{eq:Eqmn}), we can express the quantum Langevin equations (QLEs) derived based on (\ref{eq:heisen}) as follows:
\begin{align}
\dot{a}_1=& -\left(\frac{\kappa_1}{2}+\text{i}\Delta_1\right)a_1+\text{i}a_1\left(g_{11} x_{1}^{\text{th}}+g_{12} x_{2}^{\text{th}}\right) \nonumber\\ +&\text{i}J a_2+F_\text{p} \text{e}^{-\text{i}\Delta_\text{p} t}+\mathcal{N}_1,\\
\dot{a}_2=& -\left(\frac{\kappa_2}{2}+\text{i}\Delta_2\right)a_2+\text{i}a_2\left(g_{21} x_{1}^{\text{th}}+g_{22} x_{2}^{\text{th}}\right) \nonumber\\ +&\text{i}J a_1+F_\text{l}+\mathcal{N}_2,\\
\dot{x}_{1}^{\text{th}}=&-\gamma_1\left(x_{1}^{\text{th}}+\beta g_{11} a_{1}^{\dagger}a_{1}+\beta g_{12} a_{2}^{\dagger}a_{2}\right),\\
\dot{x}_{2}^{\text{th}}=&-\gamma_2\left(x_{2}^{\text{th}}+\beta g_{21} a_{1}^{\dagger}a_{1}+\beta g_{22} a_{2}^{\dagger}a_{2}\right),
\end{align}
where $\kappa_1$ ($\kappa_2$) and $\mathcal{N}_1$ ($\mathcal{N}_2$) are the decay rate and quantum noise operator associated with cavity mode one (two), respectively. It is important to note that the mean values of quantum noise, Brownian noise, and the input operator are equal to zero~\cite{qu2013}.

Furthermore, in order to solve the above non-linear QLEs, we operate with a considerably weaker probe field than the control field. Thus, we can write each operator as a sum of the mean value and the first order quantum fluctuation term, i.e.,
\begin{align}
a_1=&a_{1s}+\delta a_1,\\
a_2=&a_{2s}+\delta a_2,\\
x_{1}^{\text{th}}=&x_{1s}^{\text{th}}+\delta x_{1}^{\text{th}},\\
x_{2}^{\text{th}}=&x_{2s}^{\text{th}}+\delta x_{2}^{\text{th}}.
\end{align}
The steady-state solution of the above equations can be obtained by setting the time derivative equal to zero. This procedure results in the following set of formulas:
\begin{align}
 a_{1s}=&\frac{\text{i}J a_{2s}}{\frac{\kappa_1}{2}+\text{i}\Delta_1^\prime},\\
 a_{2s}=&\frac{\text{i}J a_{1s} +F_\text{l}}{\frac{\kappa_2}{2}+\text{i}\Delta_2^\prime},\\
 x_{1s}^{\text{th}}=&-\beta\left(g_{11}|a_{1s}|^2+g_{12}|a_{2s}|^2\right),\\
 x_{2s}^{\text{th}}=&-\beta\left(g_{21}|a_{1s}|^2+g_{22}|a_{2s}|^2\right),
\end{align}
where
\begin{equation*}
	\Delta_1^\prime=\Delta_1-g_{11}x_{1s}^{\text{th}}-g_{12}x_{2s}^{\text{th}},
\end{equation*}
and
\begin{equation*}
	\Delta_2^\prime=\Delta_2-g_{21}x_{1s}^{\text{th}}-g_{22}x_{2s}^{\text{th}},
\end{equation*}
are the effective detuning values of the cavity field. The linearized QLEs of motion can be put into the following form:
\begin{align}
\delta\dot{a}_1=& -\left(\frac{\kappa_1}{2}+\text{i}\Delta_1\right)\delta a_1+\text{i}G_{11} \delta x_{1}^{\text{th}}+\text{i}G_{12} \delta x_{2}^{\text{th}} \nonumber\\ +&\text{i}J \delta a_2+F_\text{p} \text{e}^{-\text{i}\Delta_\text{p} t},\\
\delta \dot{a}_2=& -\left(\frac{\kappa_2}{2}+\text{i}\Delta_2\right)\delta a_2+\text{i}G_{21} \delta x_{1}^{\text{th}}+\text{i}G_{22} \delta x_{2}^{\text{th}} \nonumber\\ +&\text{i}J \delta a_1+F_\text{l}\\
\delta \dot{x}_{1}^{\text{th}}=&-\gamma_1\left(x_{1}^{\text{th}}+\beta G_{11} \left[\delta a_1^{*}+\delta a_1\right]\right.\nonumber \\ +&\left.\beta G_{12} \left[\delta a_2^{*}+\delta a_2\right]\right),\\
\delta \dot{x}_{2}^{\text{th}}=&-\gamma_2\left(x_{2}^{\text{th}}+\beta G_{21} \left[\delta a_1^{*}+\delta a_1\right]\right.\nonumber \\ +&\left.\beta G_{22} \left[\delta a_2^{*}+\delta a_2\right]\right),
\end{align}
where $G_{\imath\jmath}=g_{\imath\jmath}a_{\imath s}$ describes the effective coupling rate, for $\imath \in \{1,2\}$ and $\jmath \in \{1,2\}$. Both cavity modes are assumed to have equal effective photothermal relaxation rate, i.e., $\gamma=\gamma_1=\gamma_2$. Now, we solve the above linearized equations of motion perturbatively by considering the ansatzs~\cite{boyd2020}
\begin{equation}
\delta \mathcal{O}=\sum_{n\rightarrow{\{-,+\}}}\mathcal{O}_{n}\text{e}^{\text{i}n\Delta_\text{p} t},
\end{equation}
where $\mathcal{O}=\{a_1,a_2,x_{1}^{\text{th}},x_{2}^{\text{th}}\}$. Using these ansatzs, we obtain the first order solution for the outgoing probe field as 
\begin{widetext}
\begin{equation}
a_{1-}=\frac{\mathcal{X}}{\mathcal{Y}},
\end{equation}
where the perturbative calculations lead to
\begin{align*}
\mathcal{X}=&\alpha_1 F_\text{p} \left[-1+2 \alpha_{3}^2\alpha_{1}^{*}\left(\alpha_{2}^{*}-\alpha_{2}\right)\left(G_{12}G_{21}-G_{11}G_{22}\right)^2-\text{i}\alpha_{3}\left[2\alpha_{1}^{*}\left(G_{11}^{2}+G_{12}G_{21}\right)+\left(\alpha_{2}^{*}-\alpha_2\right)\left(G_{12}G_{21}+G_{22}^2\right)\right]\right.\\&\left.+\alpha_{3}\alpha_{1}^{*}\left[\alpha_2 G_{21}-\alpha_{2}^{*}\left(2G_{12}+G_{21}\right)\right]\left(G_{11}+G_{22}\right)J_1-\alpha_2^{*}\alpha_1^{*}J_1^2 \right],\\
\mathcal{Y}=&-\left(1+\alpha_2^{*}\alpha_{1}^{*}J_1^2\right)\left(1+\alpha_1\alpha_2 J_1^2\right)+\alpha_3^2\left(G_{12}G_{21}-G_{11}G_{22}\right)^2\left(\left(\alpha_1-\alpha_1^*\right)\left(-\alpha_2^*+\alpha_2\right)+4\alpha_2^*\alpha_1\alpha_1^*\alpha_2 J_1^2\right)\\&+\alpha_3\left[\text{i}\left((\alpha_1-\alpha_1^*)\left(G_{11}^2+G_{12}G_{21}\right)-(\alpha_1^*-\alpha_2)(G_{12}G_{21}+G_{22}^2)\right)-\left(\alpha_2^*\alpha_1+\alpha_1\alpha_2\right)\left(G_{12}+G_{22}\right)J_1\right.\\&\left.+\text{i}\left[-\alpha_1\alpha_1^*\alpha_2\left(G_{11}^2+G_{12}G_{22}\right)+\alpha_2^*\left(\alpha_1\alpha_1^*\left(G_{11}^{2}G_{12}G_{21}\right)-\alpha_1\alpha_2\left(G_{12}G_{21}+G_{22}^2\right)+\alpha_1^*\alpha_2\left(G_{12}G_{21}+G_{22}^2\right)\right)\right]J_1^2\right.\\&\left.-2\alpha_2^*\alpha_1\alpha_1^*\alpha_2\left(G_{12}+G_{21}\left(G_{11}+G_{22}\right)J_1^3\right)\right],
\end{align*}
\end{widetext}
and 
\begin{align*}
\alpha_1=&\frac{1}{\frac{\kappa_1}{2}+\text{i}\left(\Delta_1-\Delta_\text{p}\right)}, \\
\alpha_2=&\frac{1}{\frac{\kappa_2}{2}+\text{i}\left(\Delta_2-\Delta_\text{p}\right)},\\
\alpha_3=&\frac{\gamma \beta}{\left(\gamma-\text{i}\Delta_\text{p}\right)}.
\end{align*}

Next, we can write the standard input-output relation for the cavity field as~\cite{walls2007}
\begin{equation}
\label{eq:eout1}
E_{\text{out}}(t)+F_\text{p} \text {e}^{-\text {i} \Delta_\text{p} t}+F_\text{l}=\sqrt{2 \kappa_1} a_1,
\end{equation}
where
\begin{equation}
\label{eq:eout2}
E_{\text {out }}(t)=E_{\text {out }}^{0}+E_{\text {out }}^{-} F_\text{p} \text {e}^{-\text {i} \Delta_\text{p} t}+E_{\text {out}}^{+} F_\text{p} \text {e}^{\text {i} \Delta_\text{p} t}.
\end{equation}
By solving~(\ref{eq:eout1}) and (\ref{eq:eout2}) simultaneously, we arrive at
\begin{equation}
E_{\text {out }}^{-}=\frac{\sqrt{2 \kappa_1} a_{1-}}{F_\text{p}}-1,
\end{equation}
or
\begin{equation}
	\label{eq:ft}
E_{\text{out}}^{-}+1=\frac{\sqrt{2 \kappa_1} a_{1-}}{F_\text{p}}=F_{\text{T}},
\end{equation}
where the relation (\ref{eq:ft}) is obtained by using the homodyne technique~\cite{walls2007}. $F_{\text{T}}$ have real and imaginary parts which are expressed by
\begin{equation}
	\label{eq:absorption}
u_\text{p}=\frac{\kappa_{1}\left(a_{1-}+a_{1-}^{*}\right)}{2 F_\text{p}},
\end{equation}
and
\begin{equation}
	\label{eq:dispersion}
v_\text{p}=\frac{\kappa_{1}\left(a_{1-}-a_{1-}^{*}\right)}{2 F_\text{p}},
\end{equation}
where $u_\text{p}$ defines absorption and $v_\text{p}$ depicts the dispersion of the probe field. Similarly, we can express the 
phase dispersion of the outgoing probe field as
\begin{equation}
\Phi_{t}\left(\Delta_\text{p}\right)=\arg \left[F_\text{T}\left(\Delta_\text{p}\right)\right],
\end{equation}
which may cause transmission group delay in the vicinity of a narrow transparency window. The transmission group delay is given by
\begin{equation}
\tau_{g}=\frac{\text{d} \Phi_{t}\left(\Delta_\text{p}\right)}{\text{d} \Delta_\text{p}}=\frac{\text{d}\left\{\arg \left[F_{T}\left(\Delta_\text{p}\right)\right]\right\}}{\text{d} \Delta_\text{p}}.
\end{equation}

The sign of $\tau_{g}$ determines the property of light, with positive and negative signs indicating slow and fast light propagation, respectively. Following the introduction of a theoretical model for the multimode optical system, its dynamical equations of motion, and a linearization of the QLEs, we present some of the key findings and discussions of the proposed system.
\section{Results and discussion}
\label{Sec:results}
The photothermal effect within the optical cavity is induced by the external probe and control fields, which give rise to nonlinearity. In this section, we discuss some of our key findings regarding the importance of the above model in various processes such as nonlinear phenomena, e.g., PTIT, Fano resonances, and slow and fast light propagation. Our numerical results are based on experimentally realized parameters~\cite{ma2020}. Furthermore, for a practical analysis of the proposed scheme, we take into account realistic parameters where it is assumed that both cavity mirrors are composed of fused-silica substrates coated for high reflectivity at operational wavelength of $\lambda=1064$ nm. A high finesse cavity with low optical loss is also considered. The finesse of the cavity is calculated using the relation $\mathcal{F}=2\pi/\tau \kappa$, where $\tau=2 \textit{L}_\textit{c}/\textit{c}$ is the time taken by the photon to escape the cavity, $\textit{L}_\textit{c}$ is the length of the cavity and $\kappa$ shows the cavity decay rate. Thus, the finesse of the cavity in the proposed system is $\mathcal{F}=6000$. Also, the quality factor has a relation with high finesse given by $\mathcal{Q}=2\mathcal{F}\textit{L}_\textit{c}/\lambda=564\times 10^6$. Therefore, we assume high finesse with good quality factor, allowing us to design photothermal cavity devices with low optical loss and wide bandwidth.
\begin{figure*}
    \centering
    \includegraphics[width=0.8\linewidth]{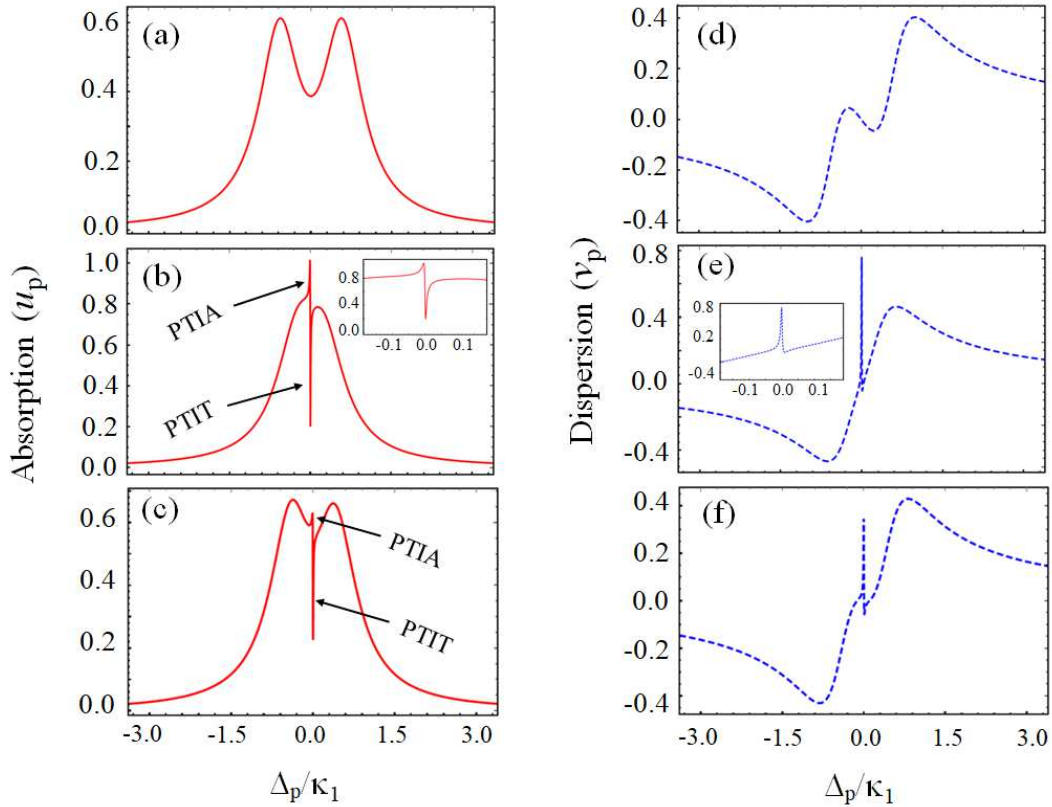}
    \caption{Absorption profile $u_\text{p}$ of $F_{\text{T}}$ versus normalized $\Delta_\text{p}/\kappa_1$ for (a) $J=0.5\kappa_1$, $\beta=0$, (b) $J=0$, $\beta=-1.8$ pm/W and $J=0.3\kappa_1$, $\beta=-1.8$ pm/W, whereas (d-f) represent their corresponding dispersion profile $v_\text{p}$. Here, $J$ signifies the coupling strength between optical modes, and $\beta$ is the effective photothermal coefficient. The insets in panels (b) and (e) represent a broader perspective of the absorption and dispersion profiles of $F_\text{T}$ at $\Delta_\text{p}\approx 0$, respectively. The other parameters are, $\kappa_1/2\pi=0.6$MHz, $\gamma/2\pi=15.4$ Hz, $\kappa_1/2\pi=530$ KHz, $\kappa_2/2\pi=430$ KHz, $\Delta_1=\Delta_2=0$, $G_{11}/2\pi=220\kappa_1$ and $G_{11}=G_{12}=G_{21}=G_{22}$. Where PTIT and PTIA stand for photothermal induced transparency and absorption, respectively.}
    \label{fig:figure02}
\end{figure*}
\subsection{Multimode photothermally-induced transparency and absorption (PTIT and PTIA)}
The EIT and EIA phenomena have been observed in coupled systems that account for radiation pressure forces~\cite{lei2015three,qu2013}; this is not the case in the proposed system, where PTIT and PTIA arise as a result of photothermal forces. The photothermal effect significantly reduces the Brownian fluctuation of mechanical oscillators~\cite{metzger2004c,Gigan2006}. Additionally, photothermal parameters such as the photothermal relaxation rate and photothermal coefficient can be defined by modulating the frequency of the input laser or expanding the photothermal effect, which results in a change in cavity length. The PTIT phenomenon is easily accessible experimentally, which opens the door to applications in traditional signal processing such as filtering and optical amplification. Another distinct characteristic of this scheme is its compact and versatile design, which allows for quick and precise characterization of photothermal effects. Furthermore, photothermal effects have been associated with a variety of other phenomena, including self-sustaining oscillations~\cite{Ludwig2008}, chaos~\cite{Carmon2007}, the production of squeezed light~\cite{goda2005}, and so on. All of these aspects contribute to the development of photothermal optics. In this subsection, we present results for transparency and absorption phenomena induced due to the photothermal effects in this multimode optical system.

We display in Fig.~\ref{fig:figure02} the absorption $u_\text{p}$ (\ref{eq:absorption}) and dispersion $v_\text{p}$ (\ref{eq:dispersion}) profile of the outgoing probe field versus normalized $\Delta_\text{p}/\kappa_1$. First, we consider the case when the two optical cavities are only coupled through hopping coefficient $J$ and have no connection with the mechanical modes because due to the absence of photothermal effects, i.e, $\beta=0$. In this case, we get only one transparency dips which arises due to the hopping coefficient $J$, see Fig. \ref{fig:figure02}(a). Consider the opposite case, where the hopping coefficient $J$ is set to zero and the photothermal coefficient is non-zero. In this scenario, the interaction of the optical cavity with the mechanical mode via photothermal effects results in a sharp transparency dip and absorption peak, as depicted in Fig. \ref{fig:figure02}(b), where PTIT and PTIA stand for photothermally-induced transparency and absorption, respectively. The inset depicts a broader perspective, disclosing that the sharp transparency dip is positioned at $\Delta_\text{p}=0$. Following that, the hopping and photothermal coefficients are both set to non-zero values, specifically $J=0.5\kappa_1$ and $\beta=-1.8$ pm/W. In this case, we see double transparency dips and single absorption peaks caused by hopping and photothermal coefficients, demonstrating that when optical cavities are connected with mechanical modes via photothermal effects and with each other, we observe double transparency dips. Therefore we would like to investigate how single to double transparency dips occur in the considered optical system when photothermal effects are taken into account. This discussion lays the groundwork for delving deeper into how the control parameters influence the amplitude and width of transparency dips, to be further explored below.
\begin{figure}
    \centering
    \includegraphics[width=1\linewidth]{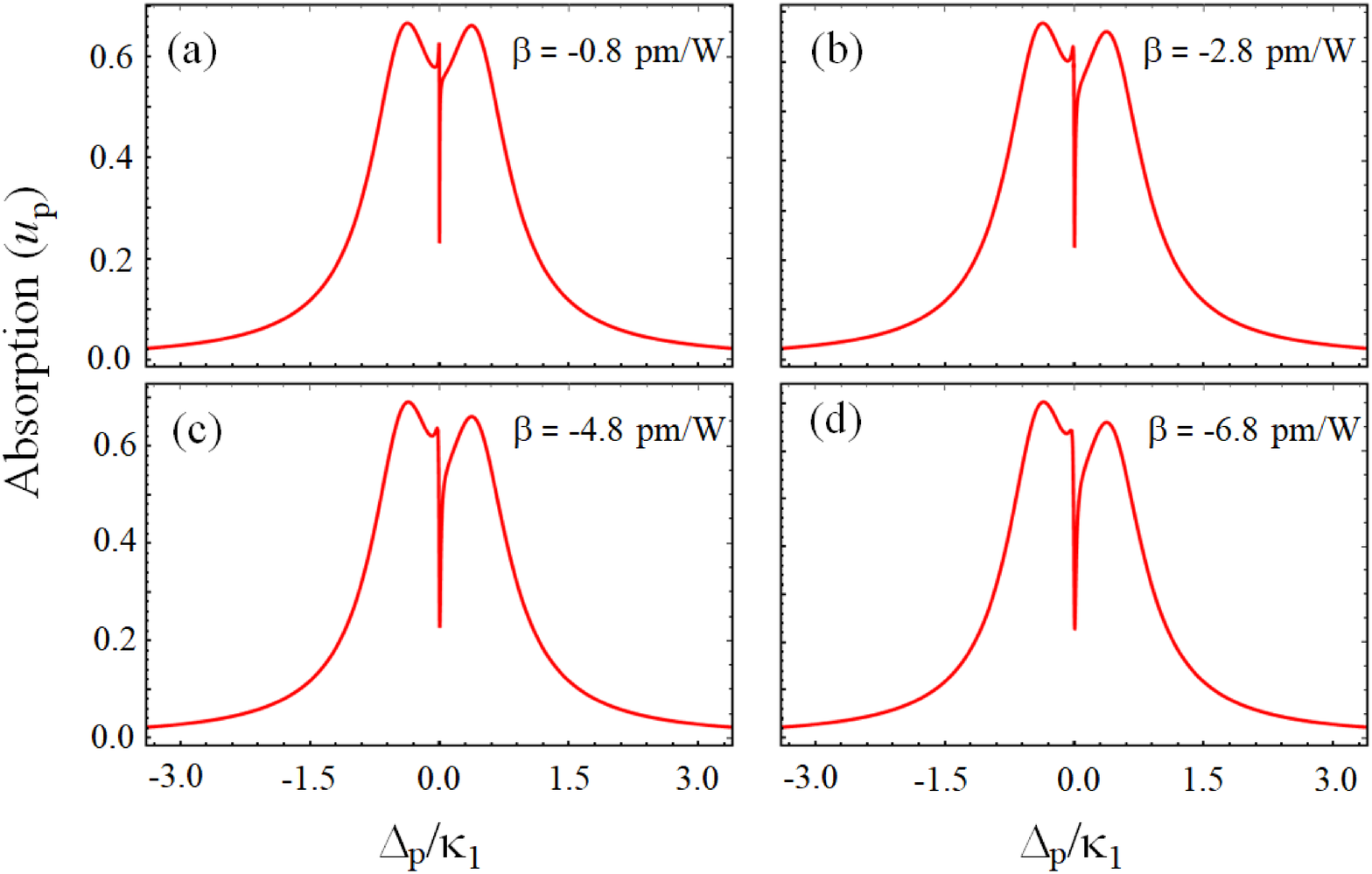}
    \caption{Absorption profile $u_\text{p}$ of $F_{\text{T}}$ versus normalized $\Delta_\text{p}/\kappa_1$ for (a) $\beta=-0.8$ pm/W, (b) $\beta=-2.8$ pm/W, (c) $\beta=-4.8$ pm/W and (d) $\beta=-6.8$ pm/W. Where $\beta$ is the photothermal coefficient. The remaining parameters are same as given in Fig.~\ref{fig:figure02}.}
    \label{fig:figure03}
\end{figure}

We turn now to an investigation of the role played by the control parameters that affect the amplitude and width of the transparency dips of this multimode optical system. First, it is highly desirable to examine how the photothermal effect, as represented by the photothermal coefficient $\beta$, impacts the transparency phenomenon. Fig. \ref{fig:figure03} (a-d) shows the absorption profile of the outgoing field plotted as a function of $\Delta_\text{p}/\kappa_1$ for different values of $\beta\in-\{0.8, 2.8, 4.8, 6.8\}$ pm/W. In each panel, we observe two sharp transparency dips and a single absorption peak. The first transparency dip is located at $\Delta_\text{p}\approx -0.1\kappa_1$, whereas the second at $\Delta_\text{p}=0$.
Moreover, increasing the value of $\beta$ reduces the amplitude of the first (from the left side) transparency dip, demonstrating that the photothermal effect suppresses the transparency phenomenon. Thus, the coupling between two optical cavities weakens as the magnitude of photothermal effects increases.
\begin{figure}
    \centering
    \includegraphics[width=1\linewidth]{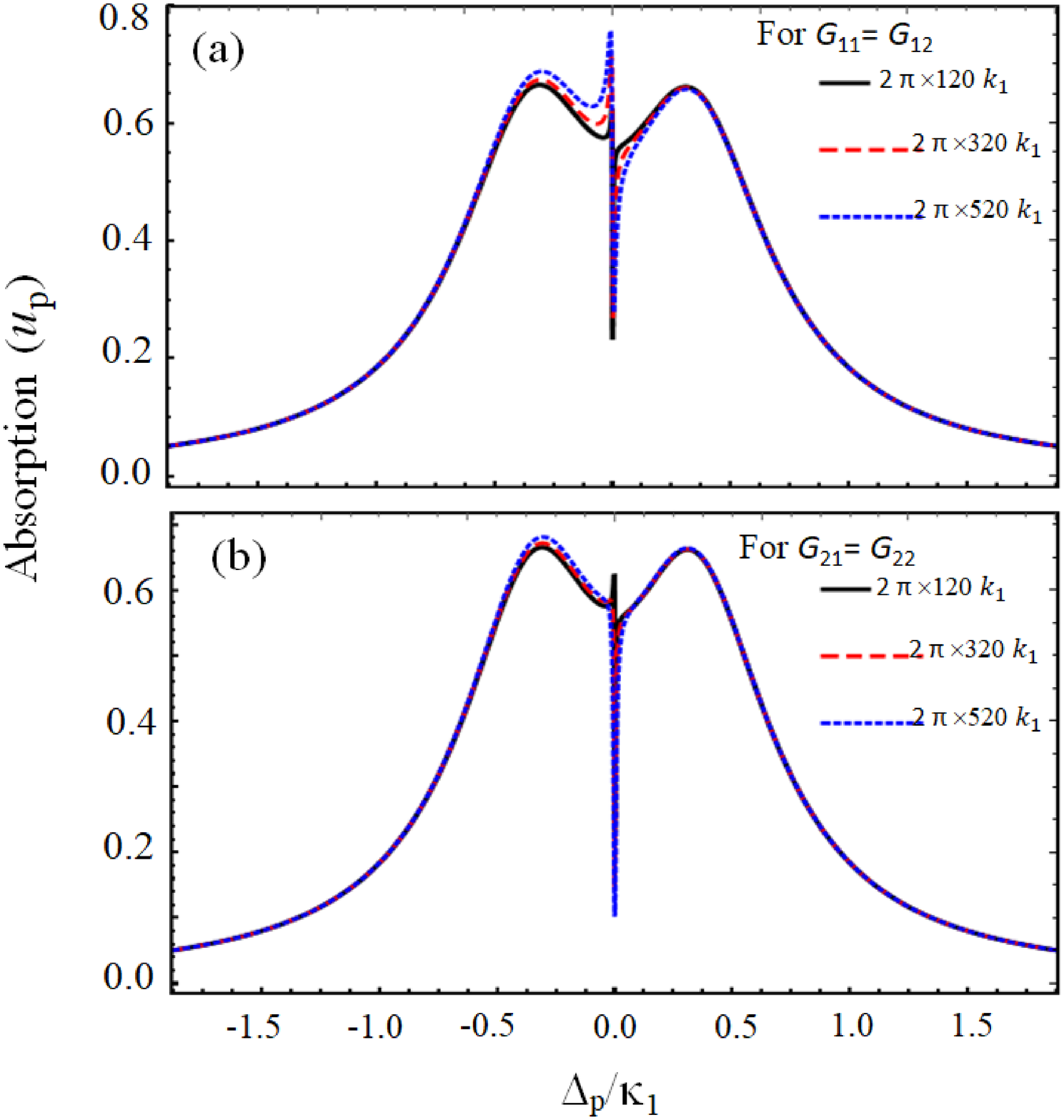}
    \caption{Absorption profile $u_\text{p}$ of $F_{\text{T}}$ versus normalized $\Delta_\text{p}/\kappa_1$ for different values of coupling rates: (a) $G_{11}=G_{12}=2\pi\times 120\kappa_1$ (black-solid), $G_{11}=G_{12}=2\pi\times 320\kappa_1$ (red-dashed) and $G_{11}=G_{12}=2\pi\times 520\kappa_1$ (blue-dotted curves); (b) $G_{21}=G_{22}=2\pi\times 120\kappa_1$ (black-solid), $G_{21}=G_{22}=2\pi\times 320\kappa_1$ (red-dashed) and $G_{21}=G_{22}=2\pi\times 520\kappa_1$ (blue-dotted curves). Where $G_{11}$, $G_{12}$, $G_{21}$ and $G_{22}$ are the coupling rates that connect the optical and mechanical modes. Rest of parameters are given in Fig.~\ref{fig:figure02}.}
    \label{fig:figure04}
\end{figure}

The absorption profile of the outgoing field can also be affected by the coupling rates between the optical cavities and mechanical modes. Fig.~\ref{fig:figure04} illustrates the absorption profile of outgoing field for different values of coupling rates. In the first case, we vary the coupling rates $G_{11}$ and $G_{12}$ while keeping the other coupling rates fixed. For $G_{11}=G_{12}=2\pi\times 120\kappa_1$ (black-solid curve), we report double transparency dips positioned at $\Delta_\text{p}\approx -0.1\kappa_1$ (left dip) and $\Delta_\text{p}=0$ (right dip), see Fig.~\ref{fig:figure04}(a). In addition, increasing the values of the coupling rates reduces the amplitudes of the both transparency dips, as described for $G_{11}=G_{12}=2\pi\times 320\kappa_1$ (red-dashed curve) and $G_{11}=G_{12}=2\pi\times 520\kappa_1$ (blue-dotted curve). Also, the sharpness and width of the second transparency dip at $\Delta_\text{p}=0$ is reduced with increase in the coupling rates, whereas the absorption peak is enhanced at $\Delta_\text{p}=0$.

In the second case, we fix the coupling rates which connect $a_1$ to $b_1$ and $b_2$, i.e., $G_{11}$ and $G_{12}$, and change the coupling rates that connect $a_2$ to $b_1$ and $b_2$, i.e., $G_{21}$ and $G_{22}$. Fig.~\ref{fig:figure04}(b) represents the absorption profile of the outgoing field for various values of coupling rates, i.e, $G_{21}$ and $G_{22}$. For $G_{21}=G_{22}=2\pi\times 120\kappa_1$, we observe double transparency dips as indicated by the black-solid curve in Fig.~\ref{fig:figure04}(b). In such case, all the coupling rates are equal to each other, i.e, $G_{11}=G_{12}=G_{21}=G_{22}=2\pi\times 120\kappa_1$. However, as $G_{21}=G_{22}$ are increased, only a single transparency dip is obtained at $\Delta_\text{p}=0$, while the transparency observed in the first case at $\Delta_\text{p}\approx-0.1\kappa_1$ vanishes, see red-dashed and blue-dotted curves in Fig.~\ref{fig:figure04}(b). Moreover, the sharpness of the transparency also increase here, which is the converse behavior with respect to the previous scenario, where the amplitude was found to decrease.

So far in this section, we have been analyzing the PTIT and PTIA phenomena in a multimode optical system with two optical modes associated with a hopping coefficient $J$ and mechanical modes connected by photothermal effects. The single to double transparency dips and a single absorption peak caused by hopping, as well as the photothermal coefficient, are illustrated. Such increases in the photothermal coefficient can reduce the transparency phenomenon. Furthermore, we examine the characteristics of outgoing field via the first cavity mode, noticing double transparency when $a_1$ interacts with mechanical modes more strongly than $a_2$. However, if $a_2$ has a stronger interaction with the mechanical modes than $a_1$, a single transparency dip can be obtained. Thus, it appears that the PTIT phenomenon can be tuned by adjusting the control parameters, namely the photothermal coefficient $\beta$, the hopping coefficient $J$, and the coupling rates $G_{\imath \jmath}$ that connect the optical and mechanical modes. In the following, we present the results for the Fano profile of the outgoing field in a multimode optical system.
\begin{figure}
    \centering
    \includegraphics[width=1\linewidth]{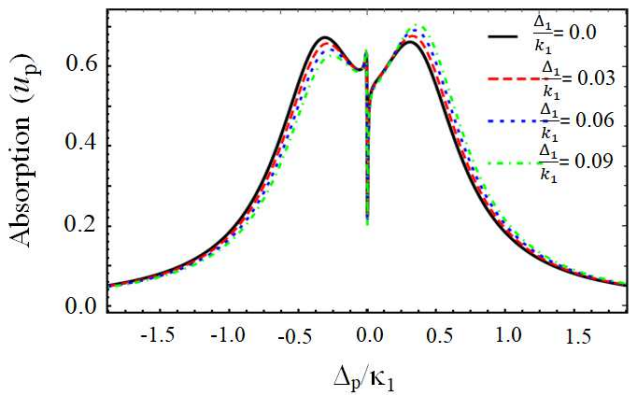}
    \caption{Fano profile in Absorption $u_\text{p}$ spectra versus normalized $\Delta_\text{p}/\kappa_1$ for different values of detuning of the cavity field $\Delta_1=0$ (black-solid), $\Delta_1=0.03\kappa_1$ (red-dashed), $\Delta_1=0.06\kappa_1$ (blue-dotted) and $\Delta_1=0.09\kappa_1$ (green-dashed-dotted curves). The detuning of the cavity field $\Delta_2=0$ remains constant. All other parameters are same as given in Fig.~\ref{fig:figure02}.}
    \label{fig:figure05}
\end{figure}
\subsection{Multimode Fano resonances}
Numerous fascinating physical phenomena, such as Fano resonances, have been observed in a variety of physical systems as a result of quantum interference between various transition pathways, which results in an absorption profile featuring minima or zero. Recently, Fano resonances have been extensively investigated theoretically in optomechanical systems, including whispering gallery modes~\cite{Vogt2017}, double cavities~\cite{qu2013}, BEC~\cite{miroshnichenko2010}, and two-level atoms (qubits)~\cite{Kamran2018}. In this subsection, we investigate Fano resonances in a multimode optical system exploiting photothermal effects, which has never been done before. Fano resonance is a nonlinear quantum interference phenomenon similar to EIT. However, it appears in an off-resonant configuration of EIT. Here, we explore the dynamics of the Fano resonances in the proposed multimode optical system by monitoring probe transmission in off-resonant cavity detuning with respect to the PTIT spectrum.  

In what follows, we examine quantitative aspects of Fano resonances induced by photothermal effects in the multimode optical system considered in this paper. Fig. \ref{fig:figure05} shows the Fano resonances in an absorption profile plotted as a function of normalized $\Delta_\text{p}/\kappa_1$ for different values of off-resonant cavity field detuning, namely, $\Delta_1=0$ (black-solid), $\Delta_1=0.03\kappa_1$ (red-dashed), $\Delta_1=0.06\kappa_1$ (blue-dotted) and $\Delta_1=0.09\kappa_1$ (green-dashed-dotted curves). At first, in the resonant condition $\Delta_1=0$, we get two symmetric absorption peaks located at $\Delta_\text{p}\approx-0.5\kappa_1$ and $\Delta_\text{p}\approx 0.5\kappa_1$, as shown by black-solid curve in Fig.~\ref{fig:figure05}. But the two peaks become asymmetric when we consider the off-resonant condition. For $\Delta_1=0.03\kappa_1$ (red-dashed curve), the absorption peak at $\Delta_\text{p}\approx 0.5\kappa_1$ have greater amplitude than the peak at $\Delta_\text{p}\approx -0.5\kappa_1$. This asymmetry behavior becomes more apparent as the cavity field detuning is increased, as illustrated in the cases of $\Delta_1=0.06\kappa_1$ (blue-dotted) and $\Delta_1=0.09\kappa_1$ (green-dashed-dotted curves). Furthermore, as the cavity detuning increases, the position of the absorption peaks shifts towards positive probe detuning ($\Delta_\text{p}>0$).
\begin{figure}
    \centering
    \includegraphics[width=1\linewidth]{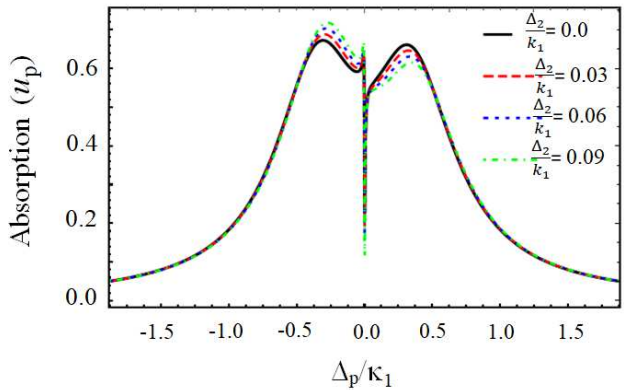}
    \caption{Fano profile in Absorption $u_\text{p}$ spectra versus normalized $\Delta_\text{p}/\kappa_1$ for different values of detuning $\Delta_2=0$ (black-solid), $\Delta_2=0.03\kappa_1$ (red-dashed), $\Delta_2=0.06\kappa_1$ (blue-dotted) and $\Delta_2=0.09\kappa_1$ (green-dashed-dotted curves). The detuning of the cavity field $\Delta_1=0$ remains constant. Remaining parameters are similar to those given in Fig.~\ref{fig:figure02}.}
    \label{fig:figure06}
\end{figure}

Following that, we look into Fano resonances in the absorption profile, first in the case where the detuning of the cavity field $\Delta_2$ changes while $\Delta_1$ remains constant. For $\Delta_2=0$ (black-solid curve in Fig.~\ref{fig:figure06}), we get two symmetric absorption peaks positioned at $\Delta_\text{p}\approx-0.5\kappa_1$ and $\Delta_\text{p}\approx 0.5\kappa_1$, which is similar to the previous case (see Fig.~\ref{fig:figure05}). However, with the off-resonance condition $\Delta_2=0.3\kappa_1$, the absorption peaks reveal an asymmetric behavior, as can be seen in the red-dashed curve in Fig.~\ref{fig:figure06}. The amplitude of the absorption peak positioned at $\Delta_\text{p}\approx-0.5\kappa_1$ is greater than that of the peak at $\Delta_\text{p}\approx0.5\kappa_1$. This asymmetric behavior can be enhanced by increasing the detuning of the cavity field, e.g., $\Delta_2=0.6\kappa_1$ (blue-dotted) and $\Delta_2=0.9\kappa_1$ (green-dashed-dotted curves) as shown in Fig.~\ref{fig:figure06}. 

Finally, we examine the Fano resonances profile in absorption spectra, which is tuned by controlling the detuning of the cavity field. First, we provide a quantitative analysis of the Fano profile by considering two different scenarios, namely when the detuning of the cavity field $\Delta_1$ changes for fixed $\Delta_2$ in the first case and vice versa in the second. Physically, in this optical system the cavity field is generated by two coherent processes. The first is the direct building up caused by the use of strong control and a weak probe fields, while the second is the scenario caused by the formation of nonlinear frequency conversion processes mediating between the optical and mechanical modes. The two paths described above provide a significant contribution to interference processes that resulted in the emergence of Fano profiles in a hybrid system. Furthermore, Fano resonances observed under off-resonant condition differ from the EIT profiles and the more typical Lorentzian resonance. In the following subsection, we investigate the fast and slow light dynamics of the group delay induced by the photothermal effects in an optical system.
\begin{figure}
    \centering
    \includegraphics[width=1\linewidth]{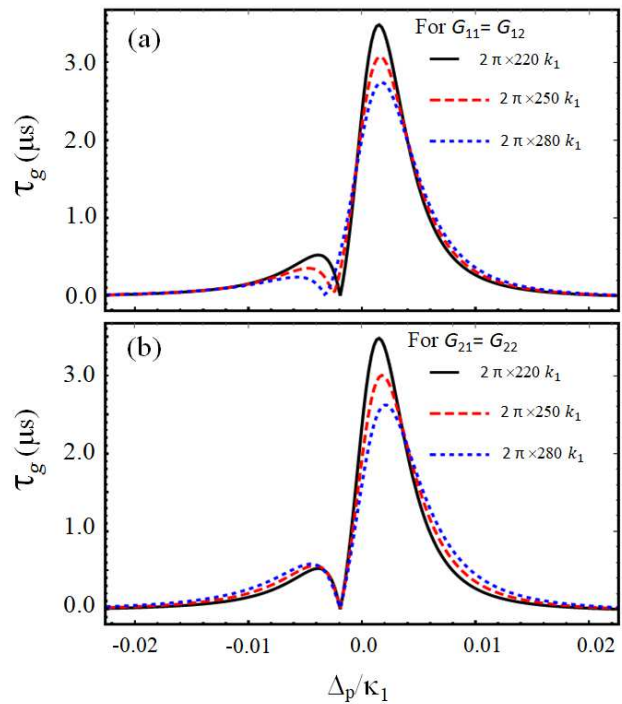}
    \caption{Dynamics of group delay $\tau_\text{g}$ versus normalized $\Delta_\text{p}/\kappa_1$ for different values of coupling rates: (a) $G_{11}=G_{12}=2\pi\times 220\kappa_1$ (black-solid), $G_{11}=G_{12}=2\pi\times 250\kappa_1$ (red-dashed) and $G_{11}=G_{12}=2\pi\times 280\kappa_1$ (blue-dotted curves); (b) $G_{21}=G_{22}=2\pi\times 220\kappa_1$ (black-solid), $G_{21}=G_{22}=2\pi\times 250\kappa_1$ (red-dashed) and $G_{21}=G_{22}=2\pi\times 280\kappa_1$ (blue-dotted curves). Where $G_{11}$, $G_{12}$, $G_{21}$ and $G_{22}$ are the coupling rates that connect the optical with mechanical modes.  Remaining parameters are similar to those given in Fig.~\ref{fig:figure02}.}
    \label{fig:figure07}
\end{figure}
\subsection{Dynamics of slow and fast light}
In the previous sections it was noted that there exists a steep dispersion behavior corresponding to PTIT at different frequencies. As the steep normal dispersion leads to group delay, steep anomalous dispersion leads to advance (fast light propagation) in the PTIT cavity system. Therefore it is constructive to study probe field transmission probability in such a setup, which has applications in optical storage and information retrieval.
In this subsection, we explore the phase response of the intracavity probe field that can give rise to group delay (slow light) or advance (fast light) in the presence of photothermal effects. In Figs.~\ref{fig:figure02} and \ref{fig:figure03}, we reported rapid phase dispersion during the transparency windows of the probe field for different values of control parameters, namely, for hopping and photothermal coefficients, resulting in a sharp decrease in the group velocity. In fact, group delay may be utilized to explain the optical response of the system to the weak probe field.

To provide a quantitative analysis, we present the results of group delay or advance in the presence of photothermal effects for different values of the coupling rates while keeping the other parameters constant. Fig.~\ref{fig:figure07} shows the dynamics of the group delay ($\tau_\text{g}$) versus normalized $\Delta_\text{p}/\kappa_1$ for different values of coupling rates, namely, $G_{11}$, $G_{12}$, $G_{21}$ and $G_{22}$. Here, it is more important to analyze the group delay in the range where the two transparency dips occur, see Fig.~\ref{fig:figure04}. First, we consider the case where the coupling rates ($G_{11}$ and $G_{12}$) that couple the cavity mode $a_1$ to the mechanical modes are changed, i.e, $G_{11}=G_{12}=2\pi\times 220\kappa_1$ (black-solid), $G_{11}=G_{12}=2\pi\times 250\kappa_1$ (red-dashed) and $G_{11}=G_{12}=2\pi\times 280\kappa_1$ (blue-dotted curves) while the rest of the coupling rates are fixed ($G_{21}=G_{22}=2\pi\times 220\kappa_1$). The group delay exhibits two amplitude peaks that equate to the two transparency dips depicted in the Fig.~\ref{fig:figure04}(a), with the first peak relating to the first transparency dip and the second peak corresponding to the second transparency dip. The sharp transparency dip in Fig.~\ref{fig:figure04}(a) produces a large amplitude peak in group delay, as shown in Fig.~\ref{fig:figure07}, indicating that we can slow down the light for a longer time. The group delay is negative at $\Delta_\text{p}\approx-0.002\kappa_1$ (PTIA peak), which is not noticeable in the figure due to scale, implying fast light. Furthermore, increasing the values of coupling rates, i.e., $G_{11}$ and $G_{12}$, reduces the group delay, as depicted in Fig.~\ref{fig:figure07}(a). Therefore, the value of group delay decreases when the optical cavity is strongly coupled with the mechanical modes.

In the second case, the coupling rates ($G_{21}$ and $G_{22}$) that couple the cavity mode $a_2$ to the mechanical modes are changed, i.e, $G_{21}=G_{22}=2\pi\times 220\kappa_1$ (black-solid), $G_{21}=G_{22}=2\pi\times 250\kappa_1$ (red-dashed) and $G_{21}=G_{22}=2\pi\times 280\kappa_1$ (blue-dotted curves) and the rest of the coupling rates are fixed ($G_{11}=G_{12}=2\pi\times 220\kappa_1$). Again, we find two amplitude peaks in the group delay (see Fig.~\ref{fig:figure07}(b)) that correspond to the two transparency dips shown in Fig.~\ref{fig:figure04}(b), and the group delay decreases as the coupling rate values increase.

In this section, we explored various physical phenomena, notably PTIT, Fano resonances, and the dynamics of slow and fast light in a multimode optical system in the presence of photothermal effects. Physically, the PTIT phenomenon is caused by destructive interference between a probe field and the anti-Stokes sideband of light scattered from a control field. Furthermore, nonlinear optical phenomena are observed in optical resonators (mechanical modes) as a result of the heating of the optical system caused by photothermal effects, which include PTIT, Fano resonances, and the dynamics of slow and fast light are briefly explored. Adjusting the controlling parameters, including the hopping and photothermal coefficients, produces single to double transparency dips. We also examine the effect of coupling rates on transparency and Fano resonances, which illustrate the strength of optical modes coupled with mechanical modes. Furthermore, the dynamics of group delay are examined in the range where the two transparency dips are observed, and it is demonstrated that a sharp transparency dip produces slow light for a longer time.

\subsection{Experimental realization of the proposed model:}
In this subsection, we present a possible experimental realization of our proposed model, along with pertinent experimental work references to facilitate the empirical analysis of such a setup. Nonreciprocal transmission across multimode systems has recently been experimentally examined using a superconducting optomechanical system with mechanical motion coupled to a multimode microwave circuit via radiation pressure~\cite{bernier2017}. Both cavity modes are linked to the same vacuum-gap capacitor in the circuit. The device operates at 200 mK in the mixing chamber of a dilution refrigerator, and all four incoming pump tones are carefully filtered and muted to remove Johnson and phase noise. One then may introduce four microwave pumps with frequencies somewhat detuned from the lower motional sidebands of the resonances in order to create a parametric coupling between the two electromagnetic and two mechanical modes. A vector network analyzer is then used to measure the reflection of an inserted probe signal around the lower (higher) frequency microwave mode. Ma et al.~\cite{ma2020} recently demonstrated by an experiment the phenomenon of transparency caused by photothermal effects in an optical cavity composed of a single cavity and a mechanical mode. Since, controlling multimode interaction is a significant task for several potential quantum system implementations in QIP, including entanglement generation~\cite{tian2013, wang2013} and quantum state transfer~\cite{Tian2012, wang2012}. Therefore, more research into photothermal effects in optical systems based on multimode interactions is highly desirable.
\section{Conclusions}
\label{Sec:conclusions}
We theoretically investigated photothermally-induced nonlinear optical phenomena in a framework where the optical field interacts with a photothermal resonator, hence leading  to photothermally-induced transparency (PTIT) and absorption (PTIA), Fano resonances, and the dynamics of slow and fast light in an optical system. Based on our analytical and numerical results, we observed single to double PTIT dips, one being sharp, and a single PTIA peak. We demonstrated how photothermal effects can suppress the transparency phenomenon by detuning the photothermal coefficient and the rates of the processes that couple the optical cavity modes to mechanical modes. We analyzed the Fano resonances profile in the presence of photothermal effects that occur in an off-resonant configuration of PTIT. Furthermore, we could achieve slow or fast light by varying the rate of the coupling between the cavity and mechanical modes.

We believe that investigating the photothermal effects that are convenient for suppressing the Brownian fluctuations of the microlever~\cite{metzger2004c,Gigan2006} and quantum ground state cooling of a mechanical resonator~\cite{pinard2008} can facilitate the development of cavity-based experiments that rely on high sensitivity.
Recent advances in the development of a photothermal effects-based all-optical switch with ultra-high tuning efficiency~\cite{Dong2022} may result in a functionally integrated component that can be employed in a broad range of efficient all-optical control applications. 
Moreover, despite the fact that photothermal forces are dissipative forces, it has been shown that a strong coupling strength can improve optomechanical entanglement~\cite{Abidi2012}. Thus, the various coupling strengths that connect cavity modes with mechanical modes in a multimode optical system via photothermal effects could help to improve cooling results and thus enhance entanglement, which has practical applications in quantum information processing.

\section*{ACKNOWLEDGMENTS}
AM acknowledges the support of postdoctoral fund of Zhejiang Normal University under Grant No. ZC304021908. GX acknowledges the support of NSF of China (Grants No. 11835011 and No. 11774316). PZ acknowledges the support of NSF of China (Grant Nos. $12174301$ and $91736104$), and the State Key Laboratory of applied optics.
%

\end{document}